\documentclass{article}

\usepackage{arxiv}

\usepackage{epsfig}

\usepackage{amssymb} 
\usepackage{amsmath} 
\usepackage{amsfonts} 
\usepackage{amsthm} 
\usepackage{empheq} 
\usepackage{mathabx} 
\usepackage{fixmath}
\usepackage{siunitx}

\usepackage{url}

\begin{document}
\title{Bidirectional recurrent neural networks for seismic event detection}
\renewcommand{\thefootnote}{\fnsymbol{footnote}} 

\author{
  Claire Birnie \\
  Equinor ASA\\
  Bergen, Norway \\
  \texttt{cebirnie@gmail.com} \\
   \And
  Fredrik Hansteen \\
  Equinor ASA\\
  Bergen, Norway \\}

\maketitle

\chead{Seismic event detection with RNNs}

\begin{abstract}
Real time, accurate passive seismic event detection is a critical safety measure across a range of monitoring applications from reservoir stability to carbon storage to volcanic tremor detection. The most common detection procedure remains the Short-Term-Average to Long-Term-Average (STA/LTA) trigger despite its common pitfalls of requiring a signal-to-noise ratio greater than one and being highly sensitive to the trigger parameters. Whilst numerous alternatives have been proposed, they often are tailored to a specific monitoring setting and therefore cannot be globally applied, or they are too computationally expensive therefore cannot be run real time. This work introduces a deep learning approach to event detection that is an alternative to the STA/LTA trigger. A bi-directional, long-short-term memory, neural network is trained solely on synthetic traces. Evaluated on synthetic and field data, the neural network approach significantly outperforms the STA/LTA trigger both on the number of correctly detected arrivals as well as on reducing the number of falsely detected events. Its real time applicability is proven with 600 traces processed in real time on a single processing unit. 

\end{abstract}

\section{Introduction}

Automated seismic event detection is the vital first step in passive seismic monitoring whether on a global or local scale. Real time, automated detection procedures play two important roles: primarily for immediate hazard detection \cite{kanamori2005}, and secondly to act as a filter prior to performing more computationally expensive procedures such as deriving the source location, magnitude and focal mechanism \cite{withers1998}. Passive seismic monitoring often focuses on microseismic events, the majority of which are observed below the noise level. Therefore, any detection procedure must be sufficiently able to detect events where the Signal to Noise Ratio (SNR) is below 1. 

One of the most common techniques for full-field event detection is the Short-Time Average to Long-Time Average (STA/LTA) method \cite{Allen1978}. However, for the STA/LTA method, and most adaptions of it, the parameters require careful selection and are closely linked to the recording conditions requiring re-calibration should the recording conditions change. Over the years many alternative approaches have been proposed for event detection, for example: waveform template matching \cite{gibbons2006}, stacking procedures \cite{Chambers2010}, and machine learning approaches \cite{zheng2018}. However, these new approaches often proved to be more computationally expensive and time-consuming than traditional detection methods (e.g., \cite{skoumal2016} and references therein) and therefore are not routinely applied. 

Recently, Machine Learning (ML) methodologies, in particular Deep Learning (DL), has seen a resurgence across all fields of seismology, and wider geoscience applications. The majority of applications have focused on the adaptation of computer vision approaches for aiding seismic interpretation. For example, salt body detection \cite{waldeland2018}, fault detection \cite{araya2017}, and horizon detection \cite{wu2018}. However, other studies have considered how ML methodologies can be utilised for seismic event detection. \cite{chen2019} proposed the combination of convolutional Neural Networks (NN) with k-means clustering for detection of seismic arrivals, whilst \cite{zhao1999} presents one of the earliest studies that considered using NNs for event detection on broadband seismometers. Novelly, they combined three different back-propagation NNs that were trained with short-, mid-, and long-term features, in a manner mimicking the short and long term fundamentals of the STA/LTA detection procedure. \cite{hochreiter1997} proposed a NN architecture called Long Short Term Memory (LSTM) that allows trends to be passed across units of the LSTM and therefore has the ability to retain information on temporal trends in the input data. Trained using data derived from rock physics experiments, \cite{zheng2018} showed that LSTM networks could be trained to develop a seismic event detection procedure with a high detection rate and reduced noise, in comparison to the STA/LTA trigger.

In this paper we advance on the study of \cite{zheng2018} by incorporating bidirectionality into the NN and training solely on synthetic data. We show that the incorporation of bidirectionality significantly improves the event detection rate, in particular reducing the number of false positives, and allows for detection of events at lower SNRs than previously possible. Training on synthetic data allowed the opportunity to create a diverse training set avoiding any possible over-engineering of the detection procedure to a specific dataset. Finally, the approach is validated on a field dataset and is shown to accurately detect the event with a low false detection rate.

\section{Data}

Supervised machine learning approaches typically require extensive labelled training datasets. These labels become the `ground truth' for the training of a machine learning model and therefore it is vital that they accurately match their data counterpart. For seismic event detection, labels indicate where a seismic arrival is present within a recording. When using recorded data for training, labels are most commonly generated by manual annotation. However, when the data has been collected as part of a  controlled experiment sometimes labels can be automatically generated through knowledge of the experimental conditions, as shown by \cite{zheng2018}. 

Two large downsides exist for using recorded data as a training dataset for passive seismic event detection. Firstly, it is impossible to determine if all events have been detected within a seismic recording; as such, the labels would only represent events that are currently detectable. Therefore, a model trained on such events would only be trained to detect events at the same detection level. Secondly, the model can only learn from the distribution it has been trained on. For example, if events were previously only detected from events with focal mechanisms resulting in positive-polarity arrivals then it is unlikely a model would detect negative-polarity arrivals as it was not exposed to these during training. To overcome these limitations, in this work we have used synthetic seismic datasets derived from a standard convolutional seismic approach. The use of synthetic seismic data for training allows the opportunity to vary the training dataset with a large range of expected event properties, as well as ensuring that no events are mislabelled in the dataset. 

\begin{figure}
  \centering
  \includegraphics[width=0.6\textwidth]{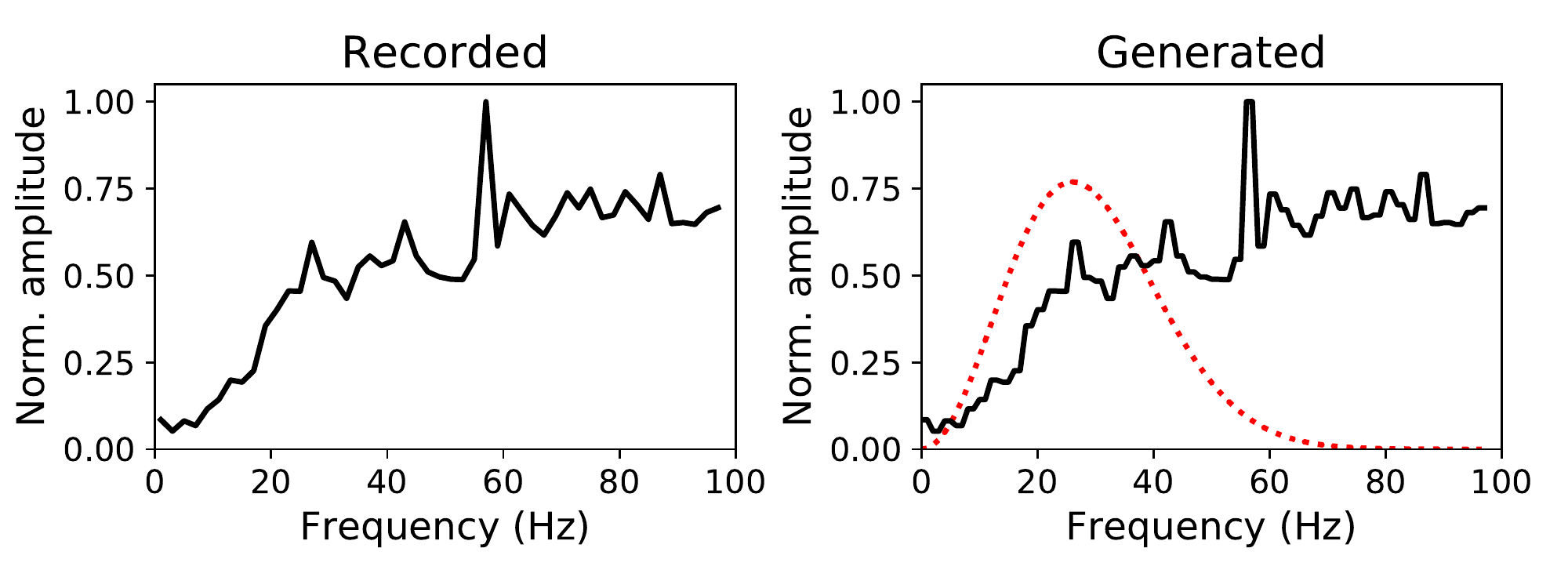}
  \caption{Frequency spectra of recorded noise (left) alongside spectra of the noise and wavelet (dashed red line) from a sample synthetic seismic trace used for model training.}
  \label{fig:spectra}
\end{figure}

Synthetic seismic traces are generated following the standard workflow for convolutional seismic traces with the input parameters and noise varying between traces. The central frequency is chosen from a uniform distribution between 20Hz and 30Hz whilst the wavelet polarity is sampled from a Bernoulli distribution with a probability of $0.5$. The event arrival time is sampled from a uniform distribution over the trace length. The SNR is sampled between $0.5$ and $2.0$ from a right-skewed Gaussian distribution to ensure a focus on low SNR events during training. The noise added to the trace is generated from many coloured Gaussian distributions scaled according to previously recorded noise such that the energy within each 2Hz window is representative of the recorded noise energy distribution, similar to the methodology proposed by \cite{Pearce1977}. Figure \ref{fig:spectra} illustrates the mean recorded energy across the frequency bands and the similar distribution observed on the generated noise models. After generation, the noise is added to the synthetic trace at the desired SNR, where the SNR is defined as the peak signal amplitude to the root-mean-square of the noise amplitudes.

\begin{figure}
  \centering
  \includegraphics[width=0.75\textwidth]{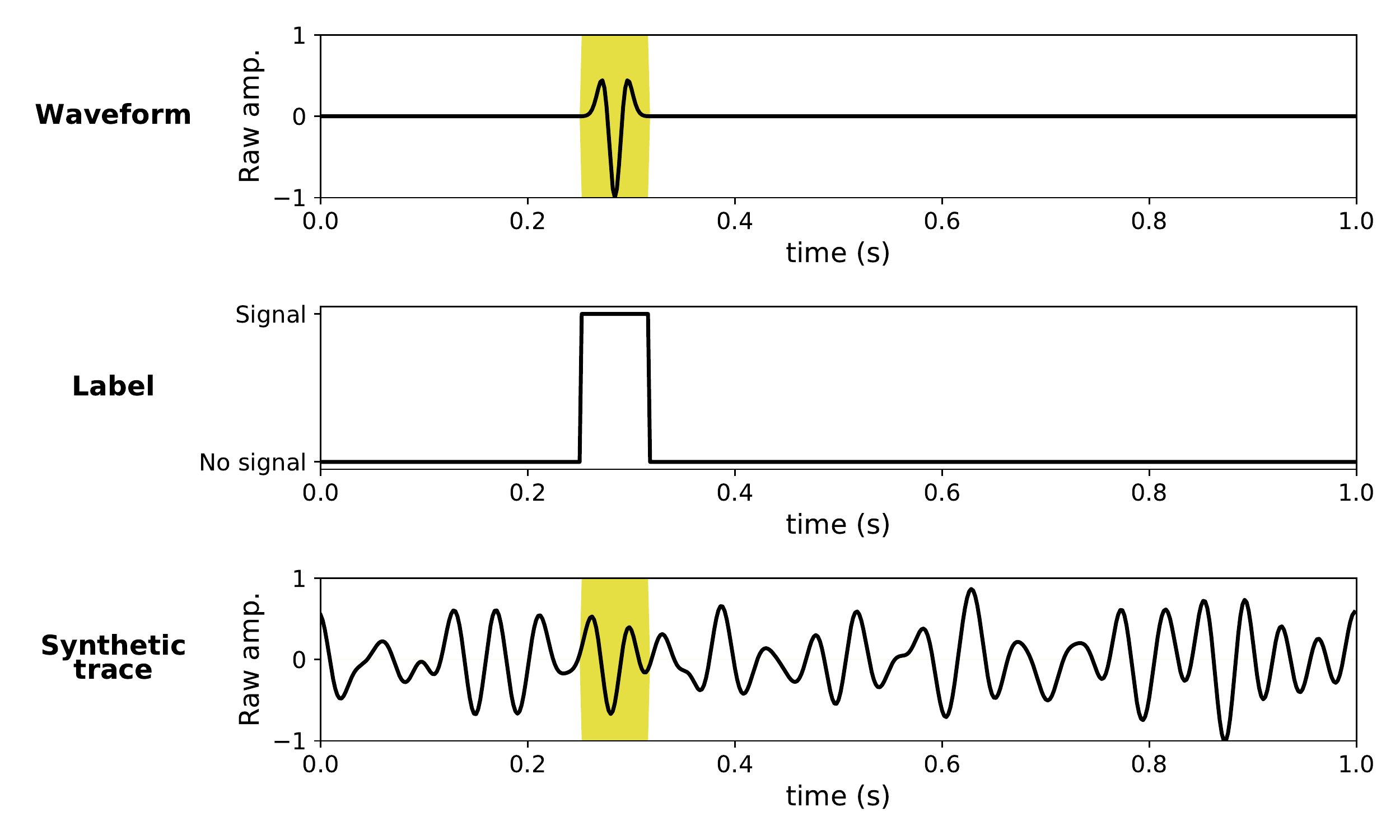}
  \caption{Labelling procedure of synthetic traces. The label (middle plot) is computed from the waveform data (top plot) prior to noise being added to generate the finalised synthetic trace (bottom plot). The highlighted area indicates where a signal is present in the normalised trace as defined by the labelling. }
  \label{fig:labelling}
\end{figure}

\begin{figure}
  \centering
  \includegraphics[width=1.\textwidth]{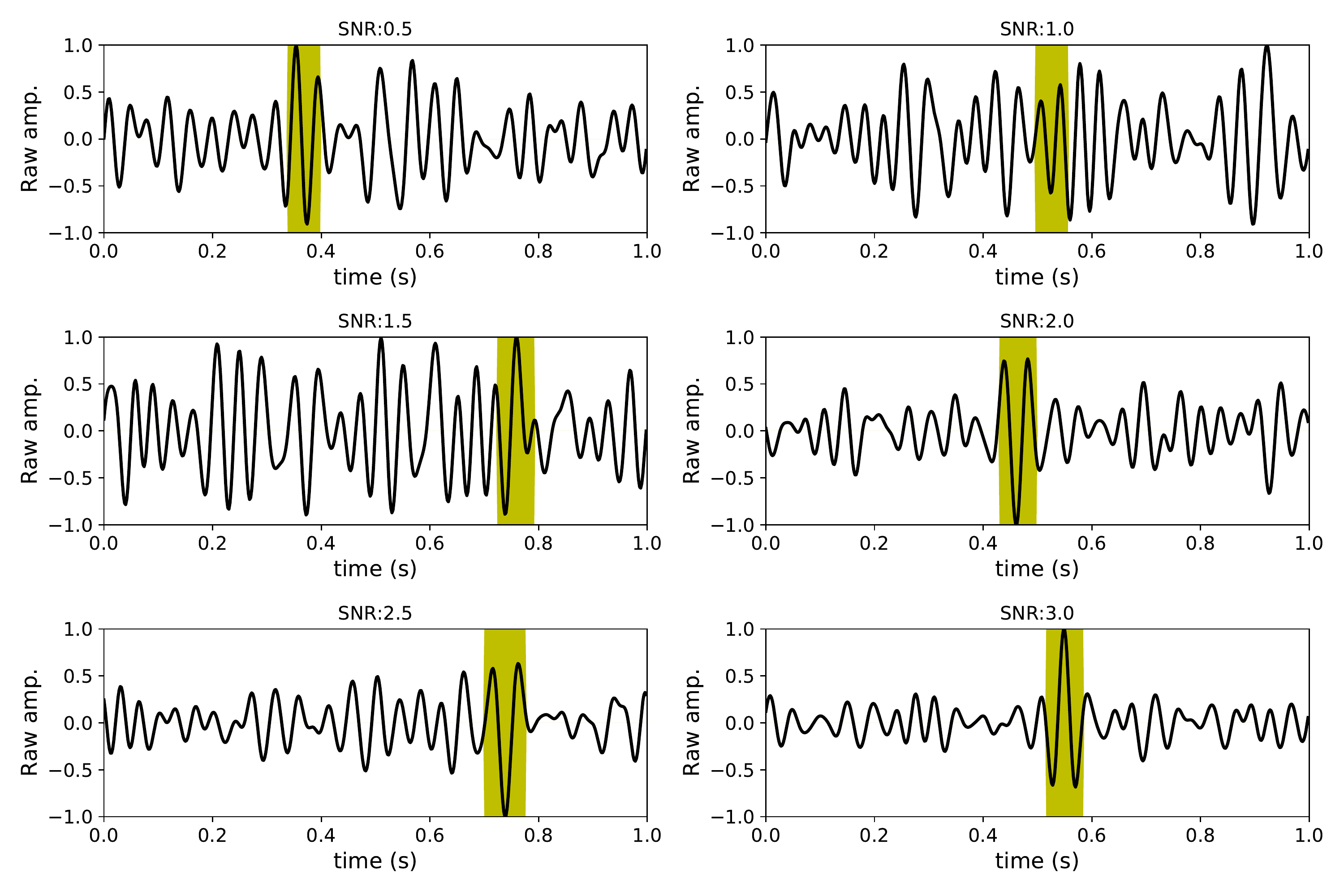}
  \caption{Examples of normalised synthetic seismic traces used for training and evaluation of event detection procedures. Highlighted area indicates where a signal is present in the trace.}
  \label{fig:synth_training_data}
\end{figure}

The labels are generated from the waveform data prior to the noise summation. An event is defined as present within the data where the absolute amplitude of the waveform is greater than \num{1e-3} as illustrated in Figure \ref{fig:labelling}. Figure \ref{fig:synth_training_data} provides examples of traces, and their respective labels, at different SNRs. The synthetic seismic traces are all bandpassed between 2 and 25Hz prior to being normalised by the maximum absolute amplitudes.

Ten thousand synthetic traces are generated for the training of the NNs, with a further 5000 synthetic traces generated for initial testing. To ensure the network does not learn to always detect an arrival, some traces contain only noise - the percentage containing only noise is considered in the gridsearch detailed below. A second test dataset is created for analysing robustness to noise. This test dataset contains 1000 traces with events at each SNR level between $0.2$ and $3.0$ with a step of $0.1$. The wavelet features and event arrival time and polarity follow the same random distributions as previously described.

To conclude, the trained networks are applied on a known event to analyse their applicability to field data. The event was recorded on a Permanent Reservoir Monitoring (PRM) system deployed on the seafloor in the North Sea. The PRM system comprises of 3458 sensors, 3C geophones with a hydrophone, arranged in a gridded-style as shown in Figure 2 of \cite{thompson2015}. This event is referred to as the ``G8 event''and has been previously analysed using a small subset of receivers by \cite{bussat2018}.

\section{Methodology}
\begin{figure}
  \centering
  \includegraphics[width=0.9\textwidth]{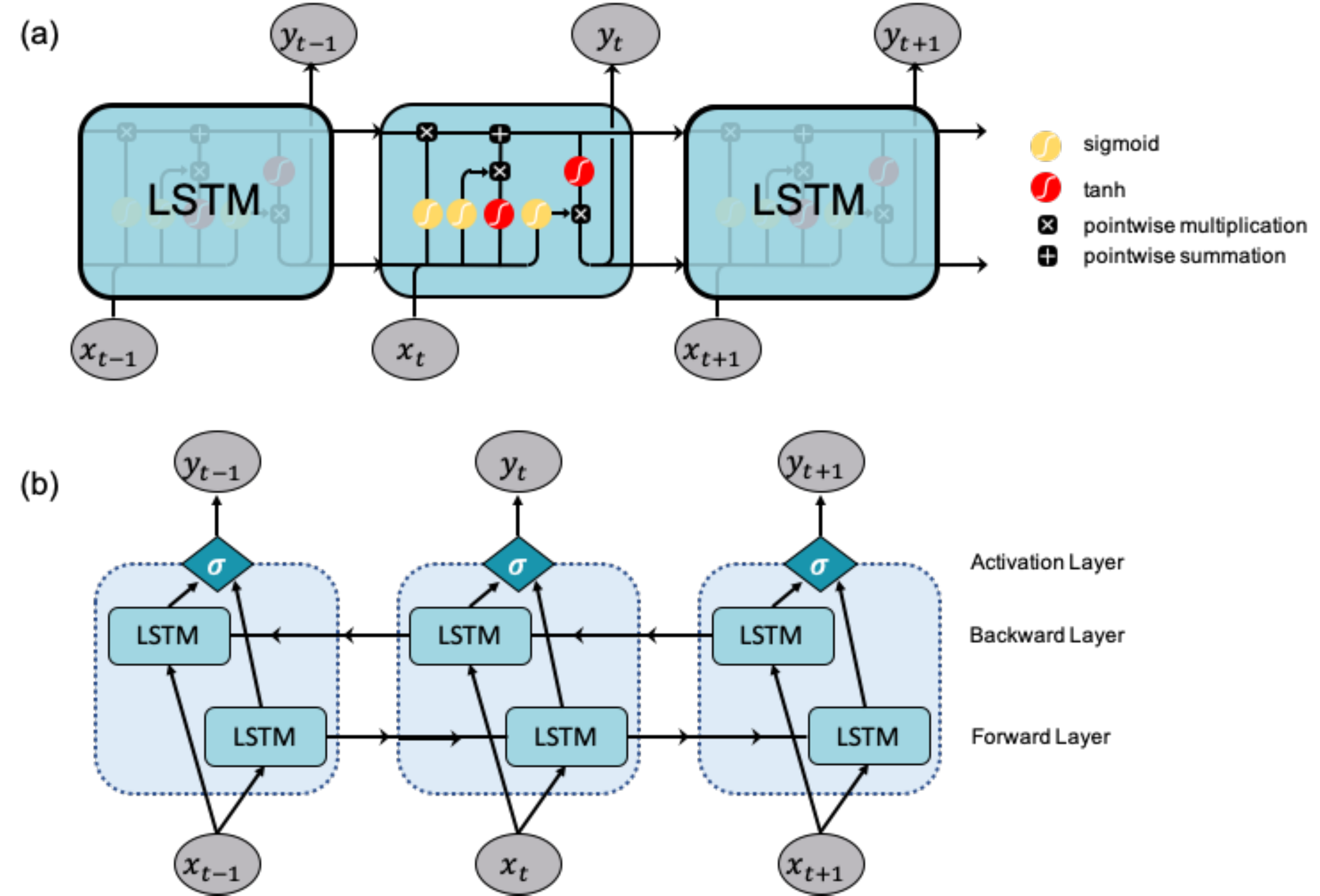}
  \caption{(a) Architecture of a single LSTM cell, and (b) architecture of a BLSTM, where $x_t$ indicates inputs and $y_t$ indicates outputs. The arrows indicate the direction of the information flow.}
  \label{fig:lstm_cell}
\end{figure}

Recurrent Neural Networks (RNN) are particularly beneficial for time series processing as they are capable of storing past inputs to produce the currently desired output, i.e., they can use earlier time steps to predict the desired outcome at a later time step. Standard RNNs are vulnerable to the vanishing gradient problem resulting in early information contributing little to the output \cite{hochreiter1998}. LSTM networks \cite{hochreiter1997} are an adaption to standard RNNs that include gates that have the ability to learn which data in a sequence is important to keep or throw away. These networks consist of units as illustrated in Figure \ref{fig:lstm_cell}(a). 

Initially proposed by \cite{kosko1988}, the inclusion of bidirectionality into NN architectures allows the network to look both forwards and backwards in time (within a given window length). This has been shown to greatly improve performance for classification and regression problems, in particular when bidirectionality is included in RNNs \cite{schuster1997}. Bi-directional LSTMs (BLSTM) consist of two layers of LSTM units, one passing information forward whilst the other passing information backwards as illustrated in Figure \ref{fig:lstm_cell}(b). 

The following section outlines the creation and training of both LSTMs and BLSTMs for the purpose of seismic event detection. As well as, the implementation of an STA/LTA autotrigger for benchmarking the NN approaches against.

\subsection{Training}
\begin{figure}
  \centering
  \includegraphics[width=1.\textwidth]{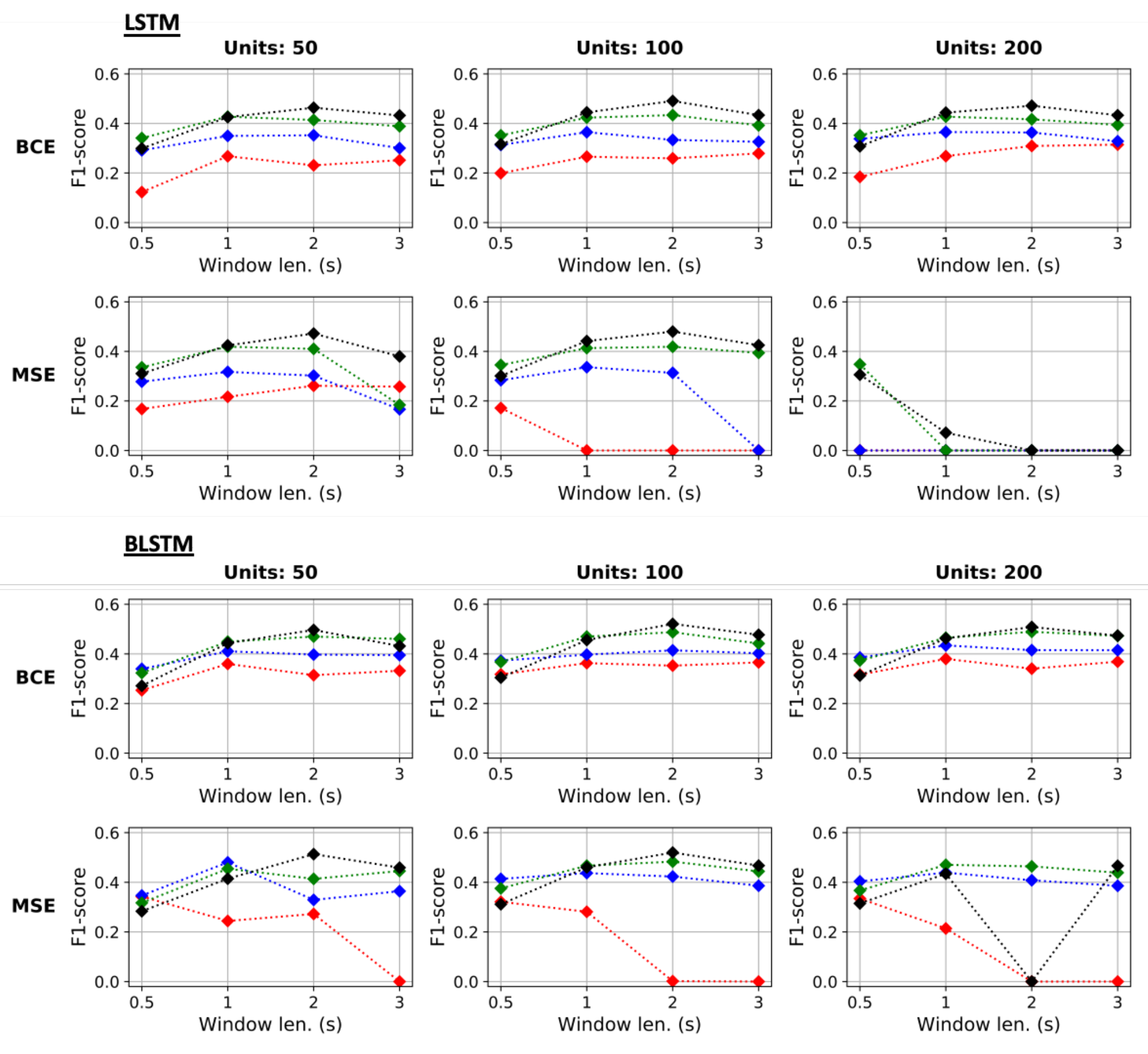}
  \caption{Grid search over window length, number of units and loss function for LSTM and BLSTM detection procedures, the top two and bottom two rows respectively. The columns indicate the numbers of units in the models. The rows preceeded by "BCE" used a binary cross-entropy loss function whilst the rows preceeded by "MSE" used a mean-squared-error loss function. The colours indicate the percent of data samples that contained a seismic event. Red lines contained 25\%, blue 50\%, green 75\% and black 100\%. }
  \label{fig:NN_gridsearch}
\end{figure}

Both the LSTM and BLSTM networks are trained with a batch size of 100, validation split of 10\% and a maximum number epochs at 100. Early stopping with a patience of 15 is employed during the training. To determine the optimum modelling parameters a grid search approach is used over the combination of parameters detailed in Table \ref{tab:NN_gridsearch}. The results of the gridsearch are illustrated in Figure \ref{fig:NN_gridsearch} where the following parameters resulted in the highest F1-score for both the LSTM and BLSTM:
\begin{itemize}
    \item \textbf{percent of traces with signal:} 100,
    \item \textbf{trace window length:} 2s,
    \item \textbf{units:} 100,
    \item \textbf{loss:} binary cross-entropy,
\end{itemize}

\begin{table}
\centering
\caption{Parameter options for LSTM and BLSTM gridsearch procedure.}
\label{tab:NN_gridsearch}
\begin{tabular}{|l|r|}
\hline
\multicolumn{1}{|c|}{\textbf{Parameter}} & \multicolumn{1}{c|}{\textbf{Options}} \\ \hline
\% of traces with event & 25, 50, 75, 100\\ \hline
Window length (s) & 0.5, 1, 2, 3\\ \hline
\# of units & 50,100,200 \\ \hline
loss function & mean squared error, binary cross entropy \\ \hline
\end{tabular}
\end{table}

\begin{figure}
  \centering
  \includegraphics[width=0.6\textwidth]{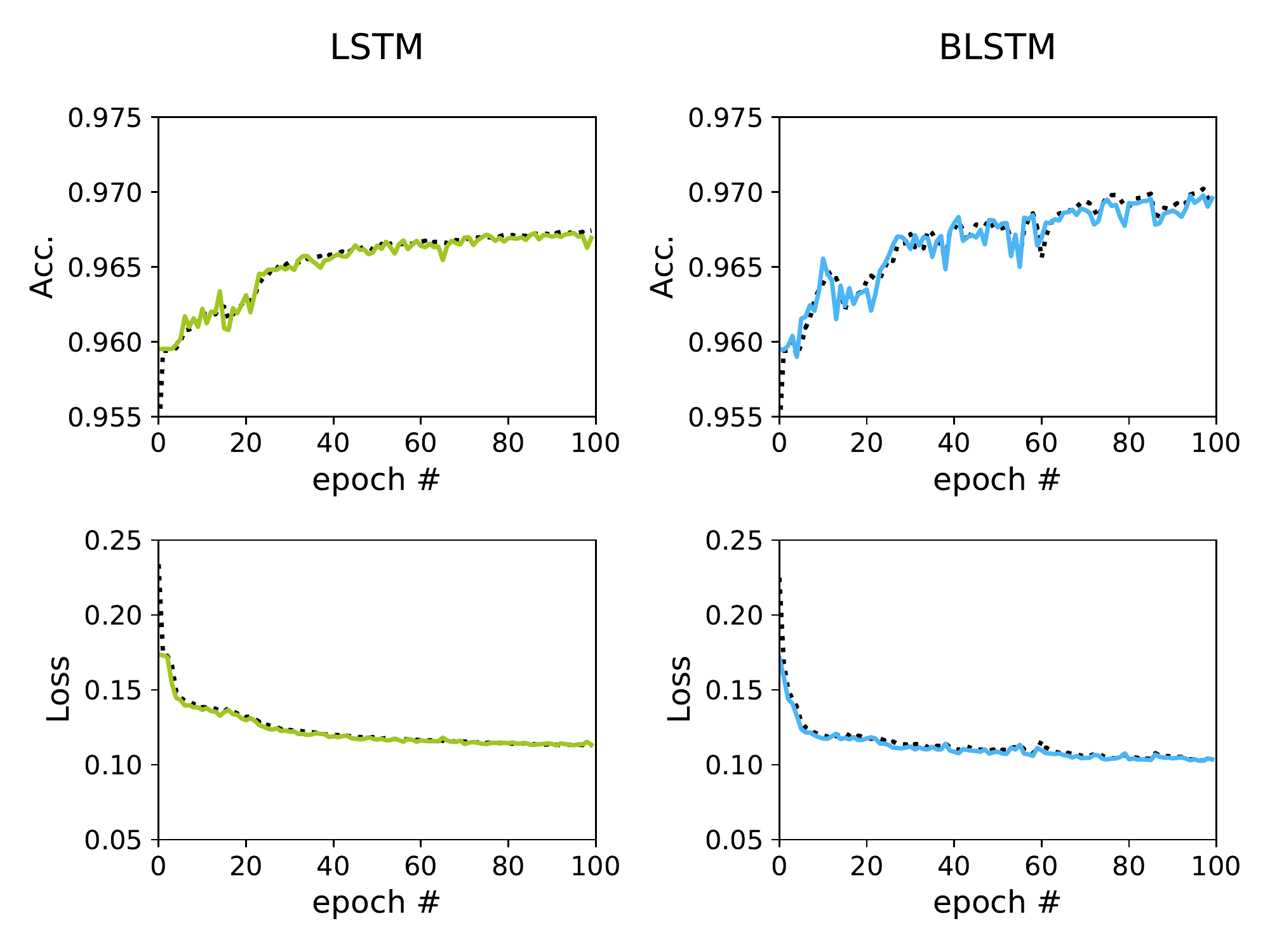}
  \caption{Evolution of accuracy and loss throughout the training of the LSTM (left) and BLSTM (left) networks. Dashed lines indicate values computed on the training data whilst solid lines indicate values calculated on the validation set.}
  \label{fig:training_loss}
\end{figure}

The models trained with these parameters are used in the analysis and benchmarking of the detection procedures detailed in the remainder of the paper. Figure \ref{fig:training_loss} illustrates the accuracy and loss as the training progresses for the models with the optimum training parameters.

\subsection{Post-processing and evaluation}
\begin{figure}
  \centering
  \includegraphics[width=0.75\textwidth]{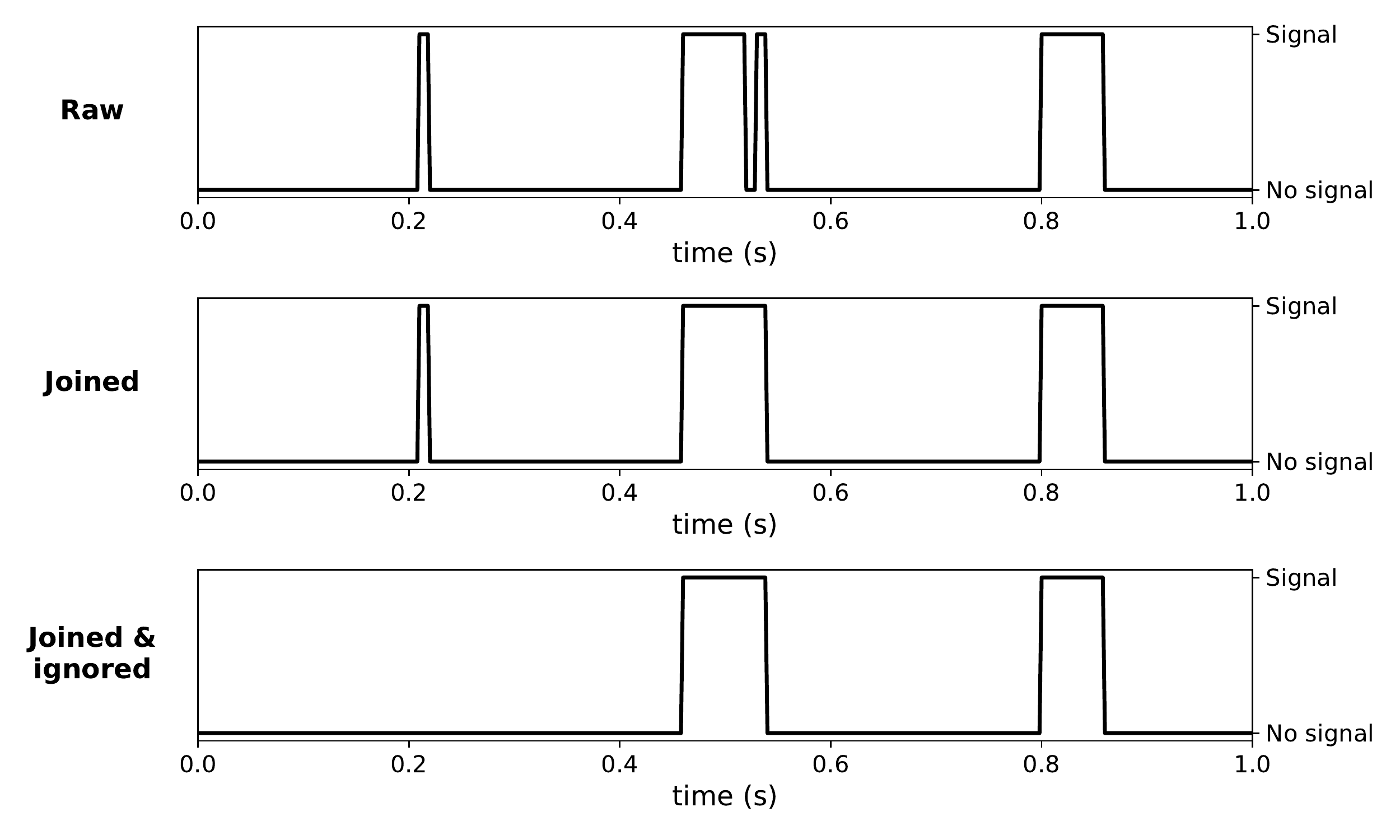}
  \caption{Post-processing of the output detection (top plot). First, nearby detected events are grouped together (middle plot), followed by a removal of short-duration events (bottom plot).}
  \label{fig:postprocessing}
\end{figure}

In the context of seismic event detection, the performance metric of a detection procedure is typically: within a time window, X, is an event accurately detected. As such, identification of the exact sample indicative of the first-break is often less important than accurate detection of an events arrival. This is slightly contradictory to most out-of-box metrics for evaluating the performance of NNs which compare sample-to-sample between a prediction and label - a resolution that is not needed in this use case. As such, a post-processing workflow is performed that groups time samples into events as outlined in Figure \ref{fig:postprocessing}. 

First detected events within 50 milliseconds of another event are grouped, where an event is defined as a number of neighbouring samples that all are identified as containing a seismic arrival. Secondly, any events with a duration of less than 25 milliseconds are removed. The resulting grouped detections are classified as seismic events.

\subsection{STA/LTA for benchmarking}
Throughout this study the STA/LTA algorithm is used for benchmarking purposes as, despite its known flaws, it is often still the default detection procedure. A classic implemenation of the STA/LTA trigger is used via the ObsPy Python package\footnote{\url{https://docs.obspy.org/packages/autogen/obspy.signal.trigger.classic_sta_lta.html#obspy.signal.trigger.classic_sta_lta}}. 

\begin{figure}
  \centering
  \includegraphics[width=1.\textwidth]{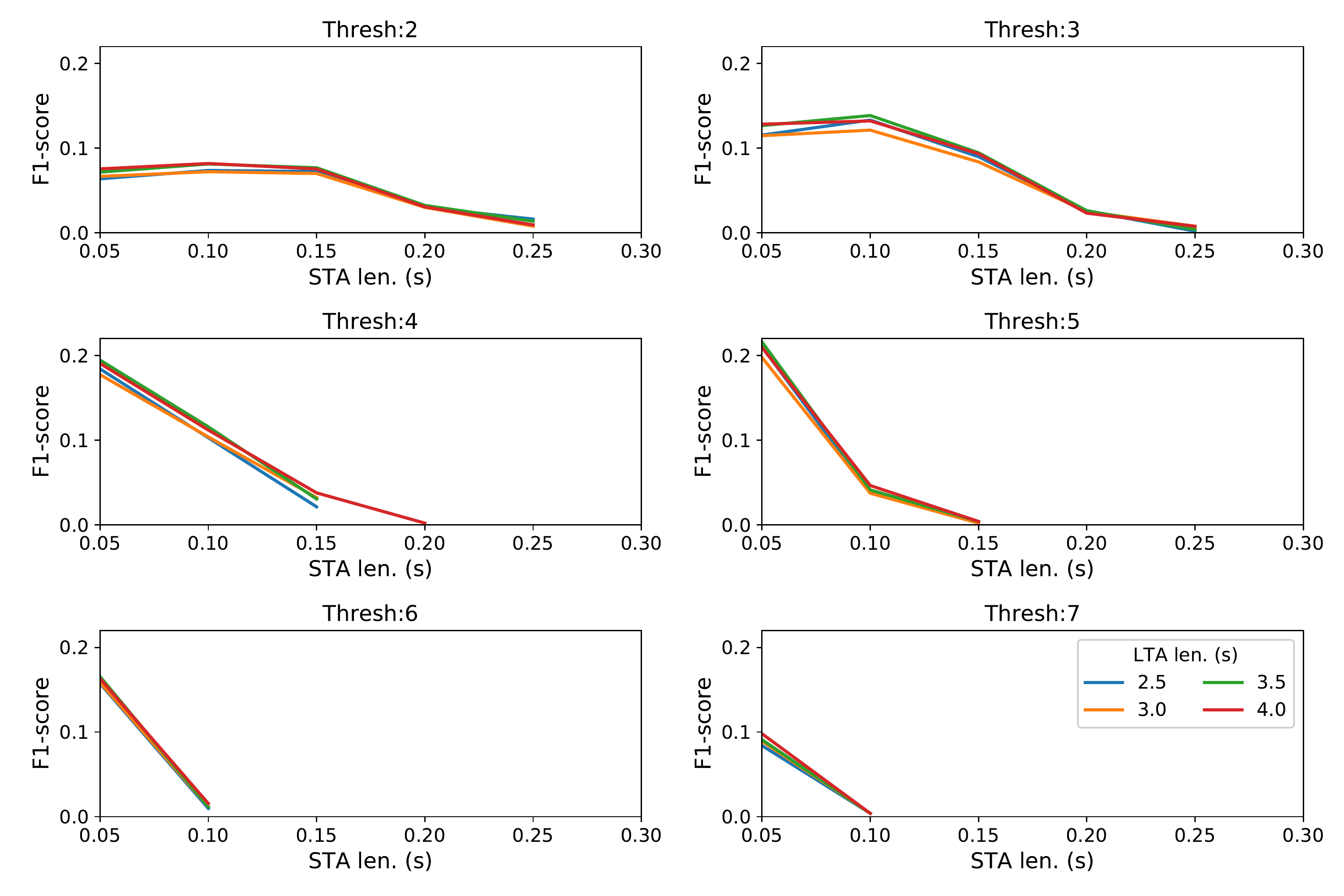}
  \caption{Grid search over STA and LTA window lengths, and detection threshold of STA/LTA ratio.}
  \label{fig:STA/LTA_gridsearch}
\end{figure}

A known downside of the STA/LTA algorithm is its sensitivity to the trigger parameters. To ensure optimum performance and a fair comparison with the ML approaches an exhaustive parameter sweep is performed across the same training datasets as used for the ML model training. Table \ref{tab:STA/LTA_par_range} details the parameter ranges investigated in the parameter sweep. The same evaluation procedure is utilised as in the ML grid search with F1-score being the evaluation metric. Figure \ref{fig:STA/LTA_gridsearch} details the performance of each parameter combination with the following parameter combination performing the best: 
\begin{itemize}
    \item \textbf{short time window:} 0.05s,
    \item \textbf{long time window:} 3.5s,
    \item \textbf{detection threshold:} 5,
\end{itemize}

\begin{table}
\centering
\caption{Parameter ranges for exhaustive parameter sweep of STA/LTA trigger performance.}
\label{tab:STA/LTA_par_range}
\begin{tabular}{|l|r|r|r|}
\hline
\textbf{Parameter} & \multicolumn{1}{l|}{\textbf{Minimum value}} & \multicolumn{1}{l|}{\textbf{Maximum value}} & \multicolumn{1}{l|}{\textbf{Step}} \\ \hline
STA win. (s) & 0.05 & 0.4 & 0.05 \\ \hline
LTA win. (s) & 1.0 & 4.0 & 0.5 \\ \hline
Threshold & 2 & 8 & 1 \\ \hline
\end{tabular}
\end{table}

This parameter combination is used for the STA/LTA trigger for benchmarking the ML model performances.

\section{Results}

\subsection{Synthetic results}
\begin{figure}
  \centering
  \includegraphics[width=1.\textwidth]{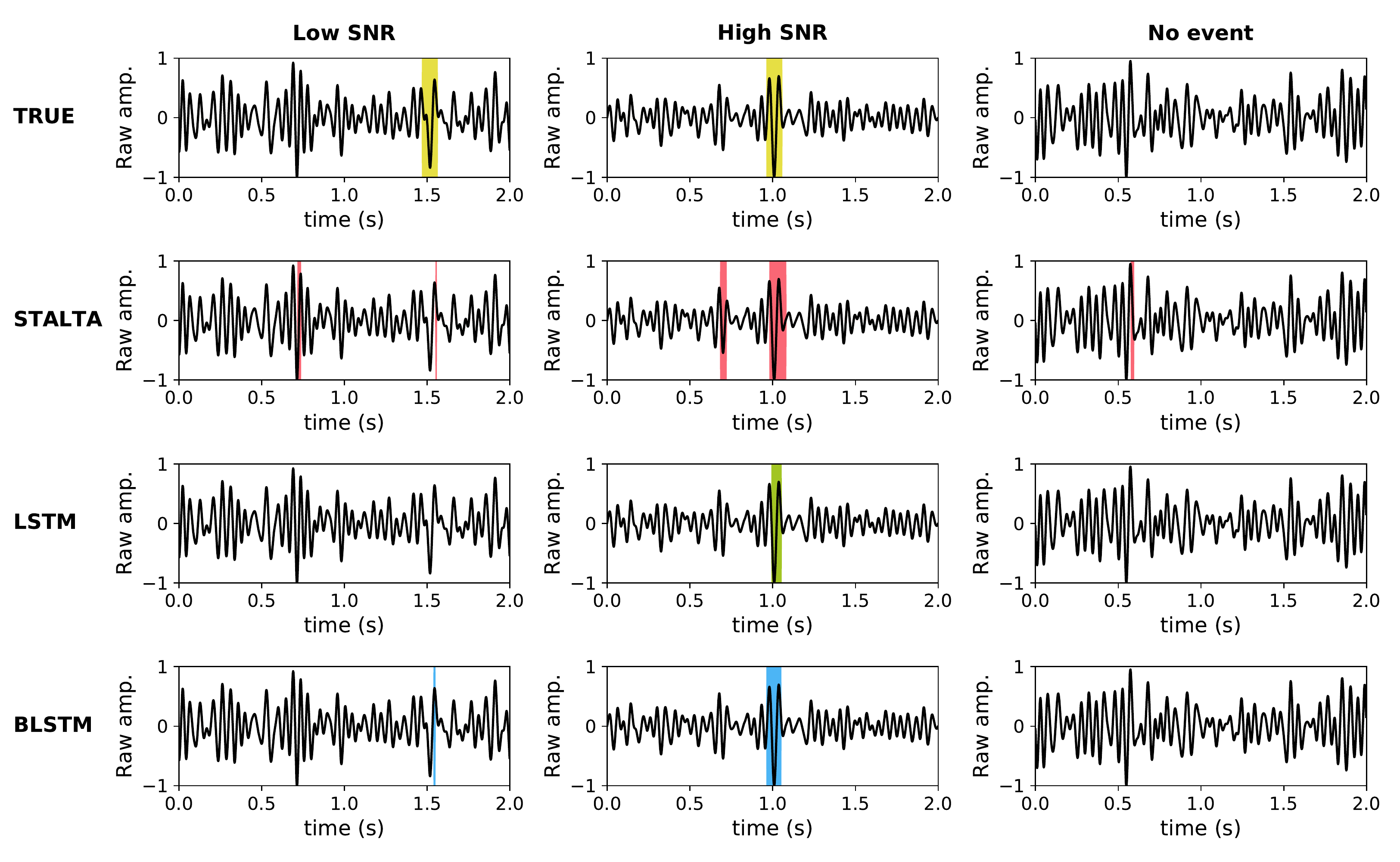}
  \caption{Event detection on synthetic seismic traces at a low SNR (left column), high SNR (middle column) and where no event is present in the trace (right column). The top row is the true label as defined when the synthetic trace was created, the following rows from top to bottom respectively represent the STA/LTA detection, LSTM detection and the BLSTM detection. Shaded areas indicate where an event was detected.}
  \label{fig:synth_detection}
\end{figure}

The STA/LTA, LSTM and BLSTM detection procedures were benchmarked against 5000 synthetic seismic traces with the characteristics as previously described. Three recording scenarios, low SNR, high SNR and no event, along with their detection results are illustrated in Figure \ref{fig:synth_detection}. The STA/LTA accurately detects both the low SNR and high SNR events, however there is also a false detection in each of the three scenarios. The LSTM accurately detects the high SNR event and creates no false detections however the low SNR event is missed. The BLSTM also creates no false detections however it successfully detects the high SNR event and a small period of the low SNR event.

\begin{figure}
  \centering
  \includegraphics[width=0.75\textwidth]{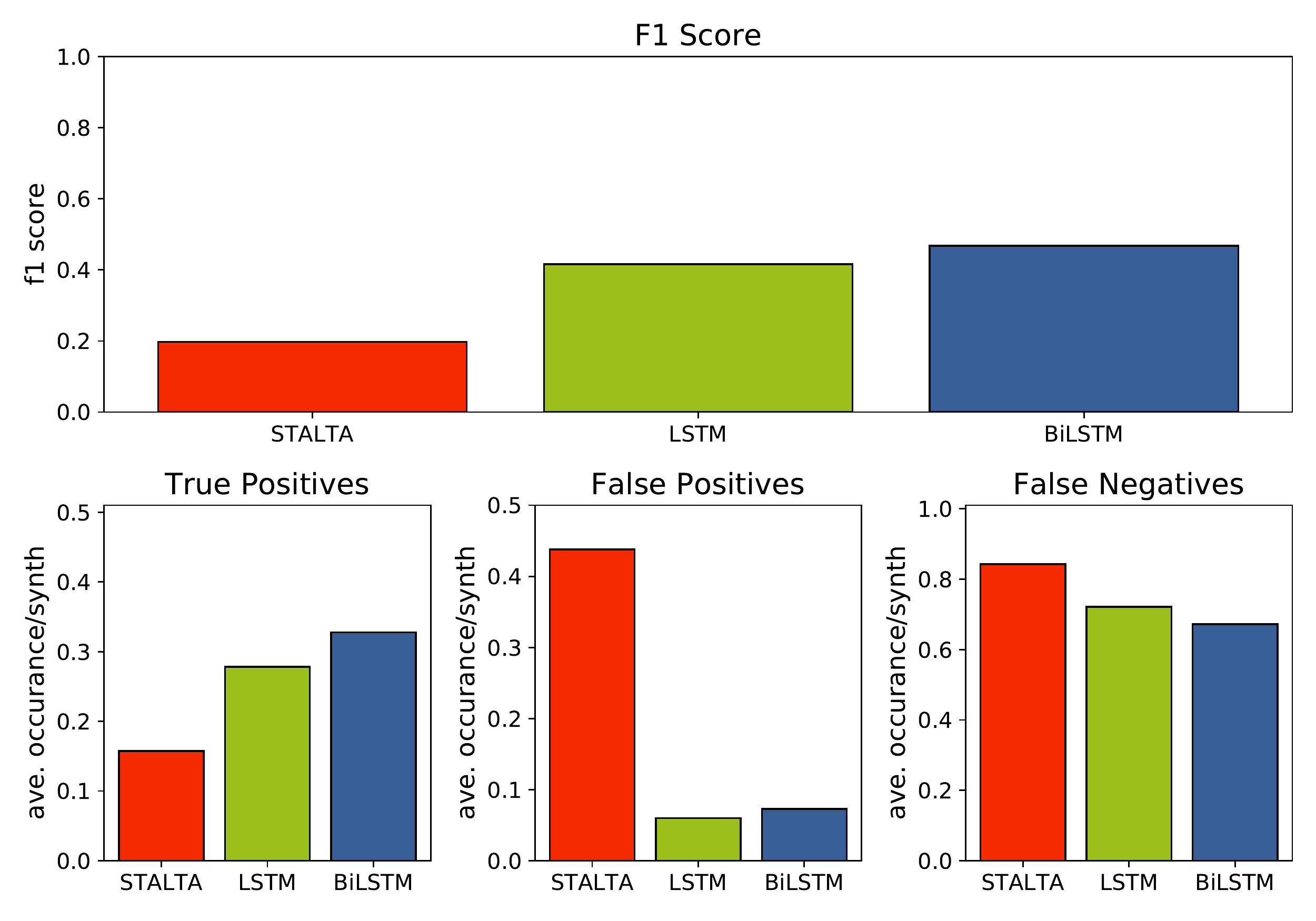}
  \caption{Evaluation of detection procedures tested on 5000 synthetic seismic traces, each containing a single seismic arrival ranging in SNR from $0.5$ to $2.0$. False positives are synonymous with false triggers whilst false negatives illustrate the number of missed events. }
  \label{fig:synth_metrics}
\end{figure}

The results for the 5000 traces as a whole are portrayed in Figure \ref{fig:synth_metrics}. The top plot highlights that overall the BLSTM outperforms both the LSTM and STA/LTA detection procedures. The bottom plots illustrate the high false positive rate arising from the STA/LTA detection procedure, whilst the BLSTM is shown to outperform LSTM the number of correctly identified events (i.e., the true positive rate).

\begin{figure}
  \centering
  \includegraphics[width=0.9\textwidth]{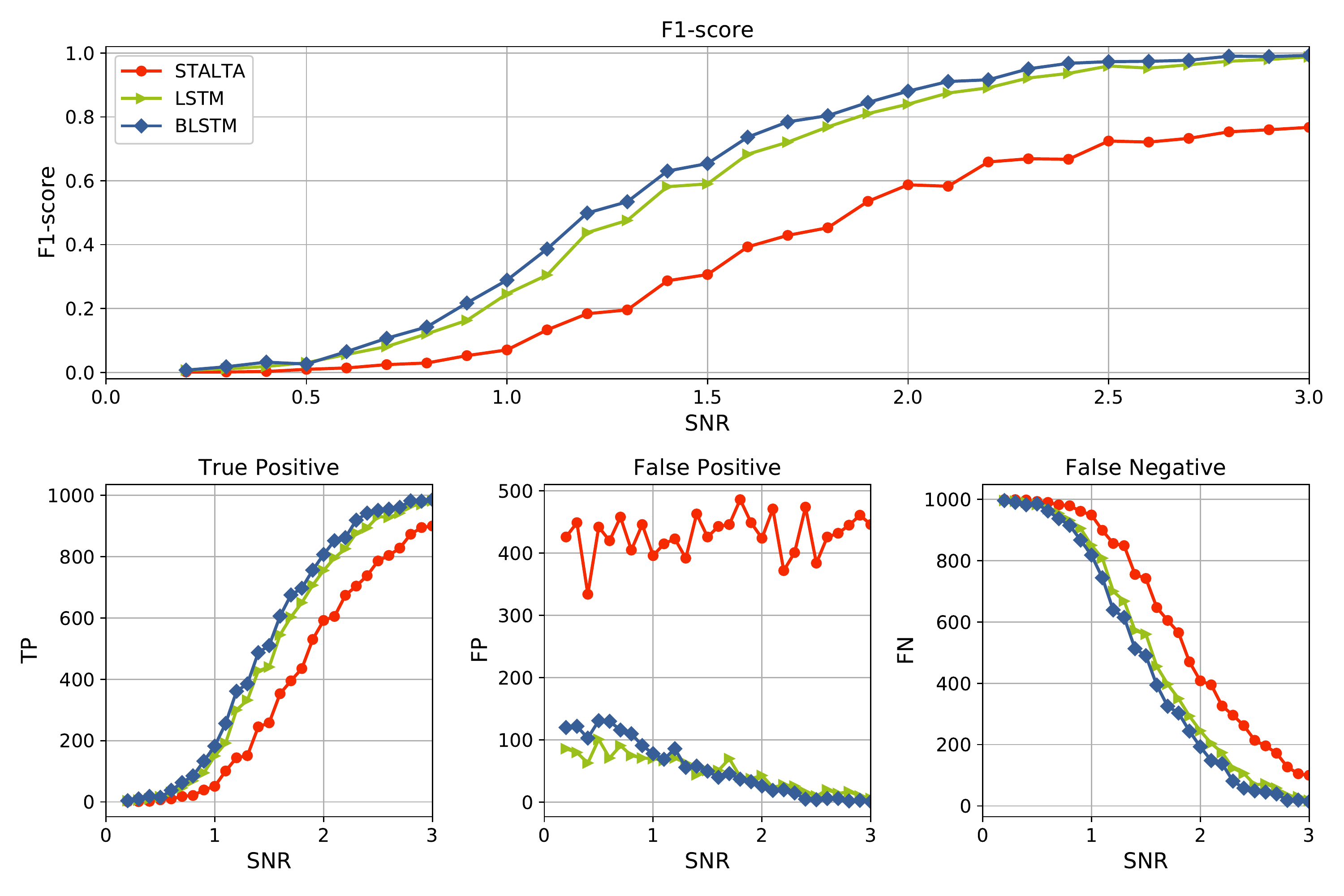}
  \caption{Analysis of the effect of noise level on the different detection procedures. The top plot illustrates the effect on the f1-score of the detections whilst the bottom three plots illustrate the number of true positives, false positives and false negatives, respectively. False positives are synonymous with false triggers whilst false negatives illustrate the number of missed events.}
  \label{fig:snr_analysis}
\end{figure}

Figure \ref{fig:snr_analysis} details how the different detection procedures perform at different SNRs. For our definition of SNR, the probability of detecting an event below an SNR of $1$ is minimal, $<20\%$, for all detection procedures, extending to $100\%$ for the NN procedures at an SNR of $2.9$. At lower SNRs ($<0.5$) there is little differentiation between the different detection procedures. Above SNR of $0.5$, the LSTM begins to outperform the STA/LTA detection procedure whilst the BLSTM begins to outperform them both around SNR of $0.7$. Above SNR of $2.7$, the performance difference between the BLSTM and LSTM detection procedures is negligible.

\subsection{Detection on field data}
\begin{figure}
  \centering
  \includegraphics[width=1.\textwidth]{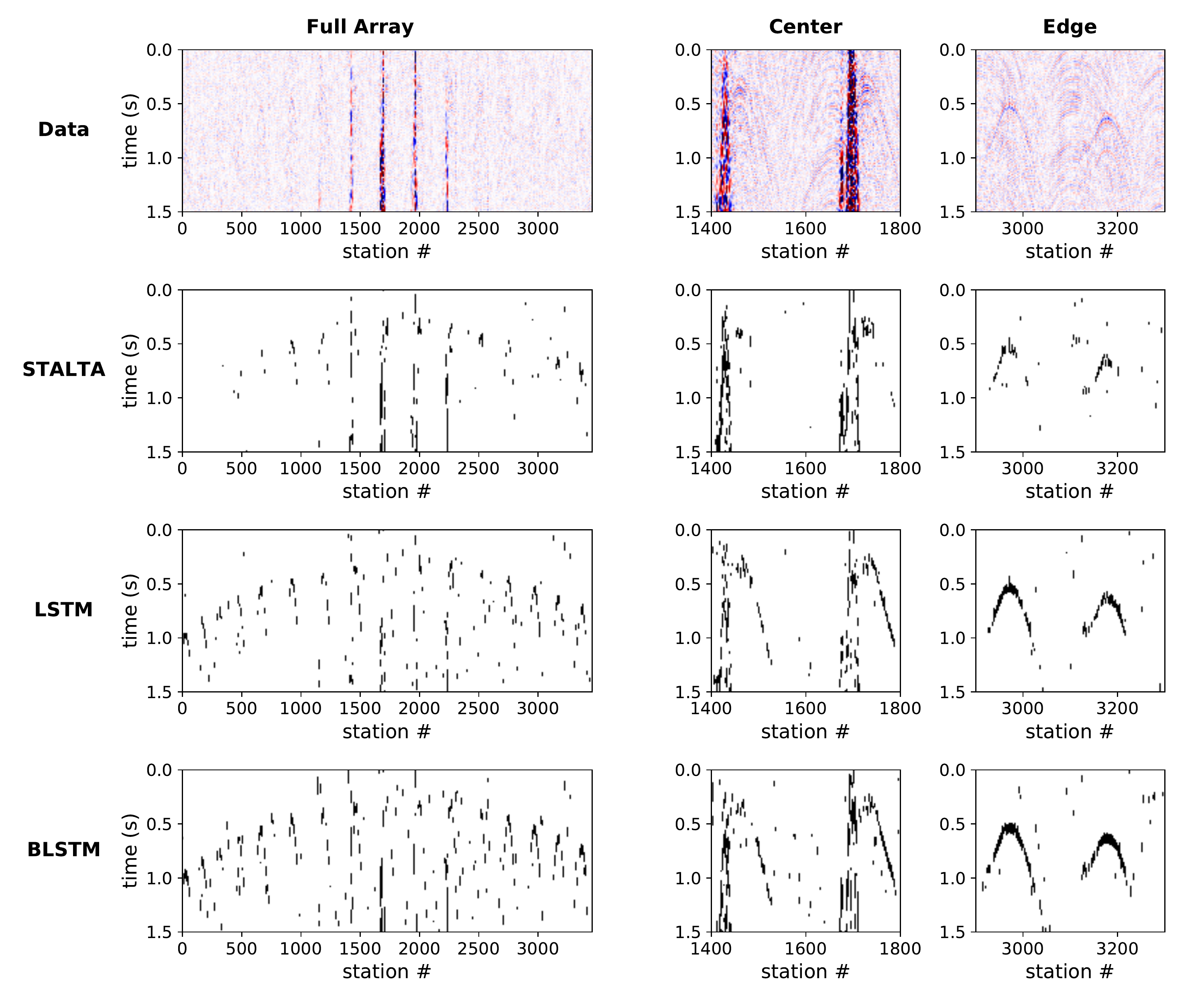}
  \caption{Field data (top) and detection results for a seismic event previously analysed by \cite{bussat2018}. The center of the array, middle column, is contaminated by platform noise, while the edge of the array has a lower noise level but is further from the event source. }
  \label{fig:G8}
\end{figure}

\begin{figure}
  \centering
  \includegraphics[width=0.8\textwidth]{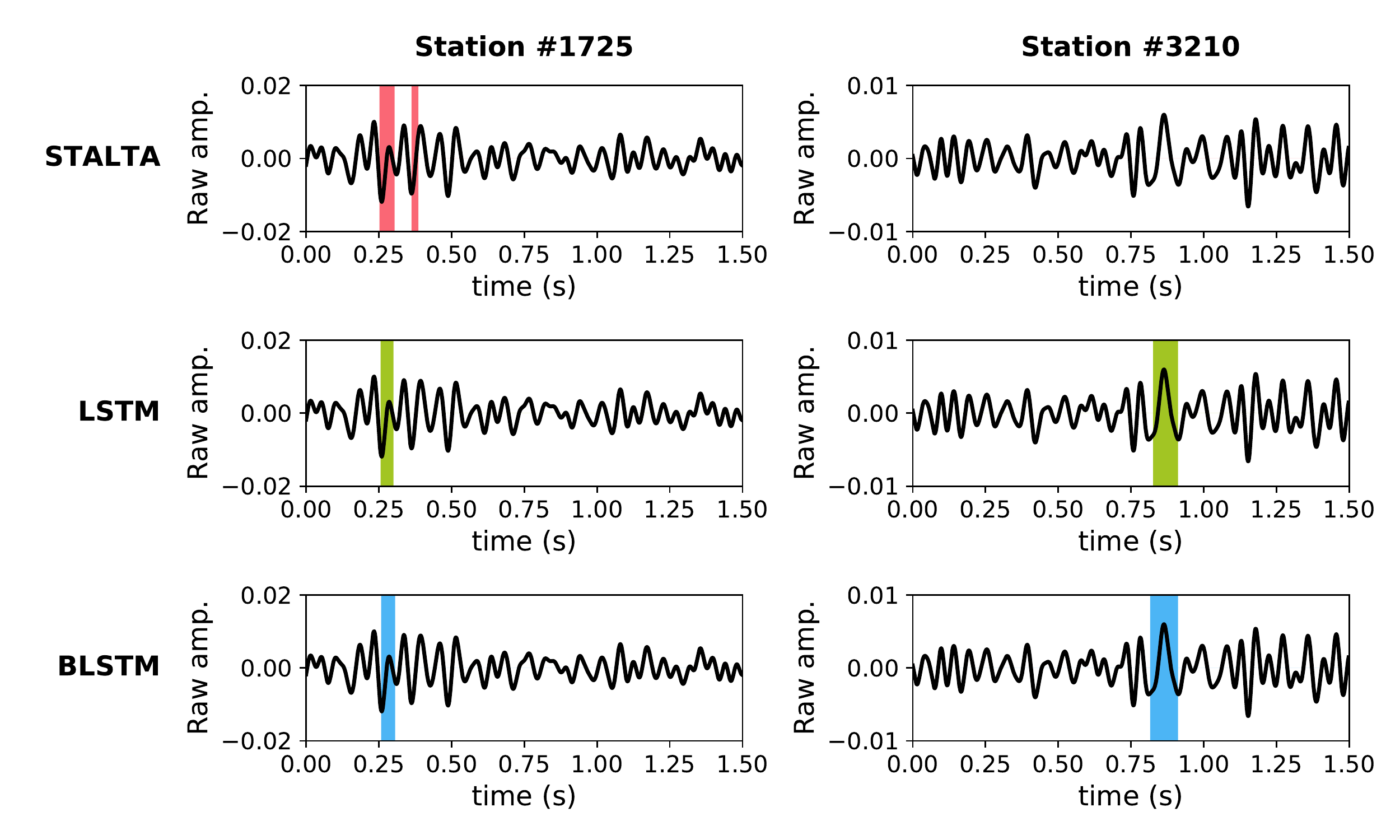}
  \caption{Detection on individual traces of G8 event from a station at the center of the array (left column) and from a station at the edge of the array (right column). Areas highlighted indicate the times when an event was detected by the STA/LTA autotrigger (top), the LSTM network (middle), and the BLSTM network (bottom).}
  \label{fig:G8_traces}
\end{figure}

Finally, the study concludes by testing all three detection procedures against a previously detected event from the Grane PRM system \cite{bussat2018}. The top row of Figure \ref{fig:G8} illustrates the z-component of the seismic data with zoomed-in sections around the noisy center of the array and the quieter edge of the array. The following three rows show the event detection results from the three detection procedures: STA/LTA, LSTM, and BLSTM respectively. Single traces with detections are illustrated in Figure \ref{fig:G8_traces}. The BLSTM detection outperforms both the STA/LTA and LSTM detections with the arrivals clearly detected on the majority of traces. However, whilst the number of false detections is reduced, they are still present within the BLSTM results. Unfortunately, all detection procedures mislabel the platform noise in the centre of the array as a seismic event.

\section{Discussion}
The three main criteria for the development and deployment of new seismic event detection procedures are:
\begin{itemize}
    \item \textbf{Automatable}: requiring no parameter adaption once in place,
    \item \textbf{Reliable at low SNRs}: events must be accurately detected below the noise level, and
    \item \textbf{Real time}: compute must be performed faster than data streaming.
\end{itemize}

The non-trivial nature of STA/LTA parameter setting and the required updating of parameters as the noise environment changes is well documented \cite{trnkoczy2012}. It was further highlighted in this study where a parameter sweep across realistic synthetic data was performed to derive the optimum STA/LTA parameters. The poor performance of the STA/LTA on the G8 event, despite its high SNR, further illustrates the non-trivial nature of parameter setting. Once trained neither the LSTM or BLSTM approaches require any parameter tuning. Assuming that the training dataset adequately captures all the different monitoring settings, the NN models can be considered as globally applicable to all monitoring settings ranging from cap-rock monitoring to seismic tremor detection. Ensuring a large variability in the training dataset is an advantage that arises from using synthetically generated data for training. However, the suitability of using synthetic traces for training is reliant on the synthetic dataset closely representing field data - a complex topic as highlighted by \cite{birnie2020}.

The STA/LTA trigger is well-known to struggle at detecting events below the noise level \cite{withers1998}. Both the LSTM and BLSTM approaches showed significant improvements on detection rates at all SNRs investigated: SNRs of $0.2 - 3.0$. Again, the use of synthetic datasets provided an advantage here over the use of field or laboratory data, allowing us the opportunity to define the SNR distribution of the training set to tailor the models to detect events at low SNRs. Alternatives to the STA/LTA such as waveform template matching or stacking procedures have much higher resilience to noise and can detect events in low SNR situations. Waveform template matching procedures, for instance, have been shown to detect events at low SNRs  and have been successfully applied for microseismic monitoring of geothermal \cite{plenkers2013} and gas \cite{song2010} reservoirs. However, they require a-priori information for the waveform template, often extracted from an event catalogue, and therefore are not easily generalisable \cite{yoon2015}. Whereas, stacking procedures typically require multiple nearby stations and are therefore not applicable for single station or sparse array applications, ruling out their use for a significant portion of hazard monitoring scenarios. Alternative approaches could be to include a noise suppression procedure prior to detection, however that would introduce an unwanted additional computational cost and due to the varying nature of noise signals it would be difficult to determine an automated noise suppression algorithm which is optimal \cite{Birnie2016,Birnie2017}. 

\begin{figure}
  \centering
  \includegraphics[width=0.5\textwidth]{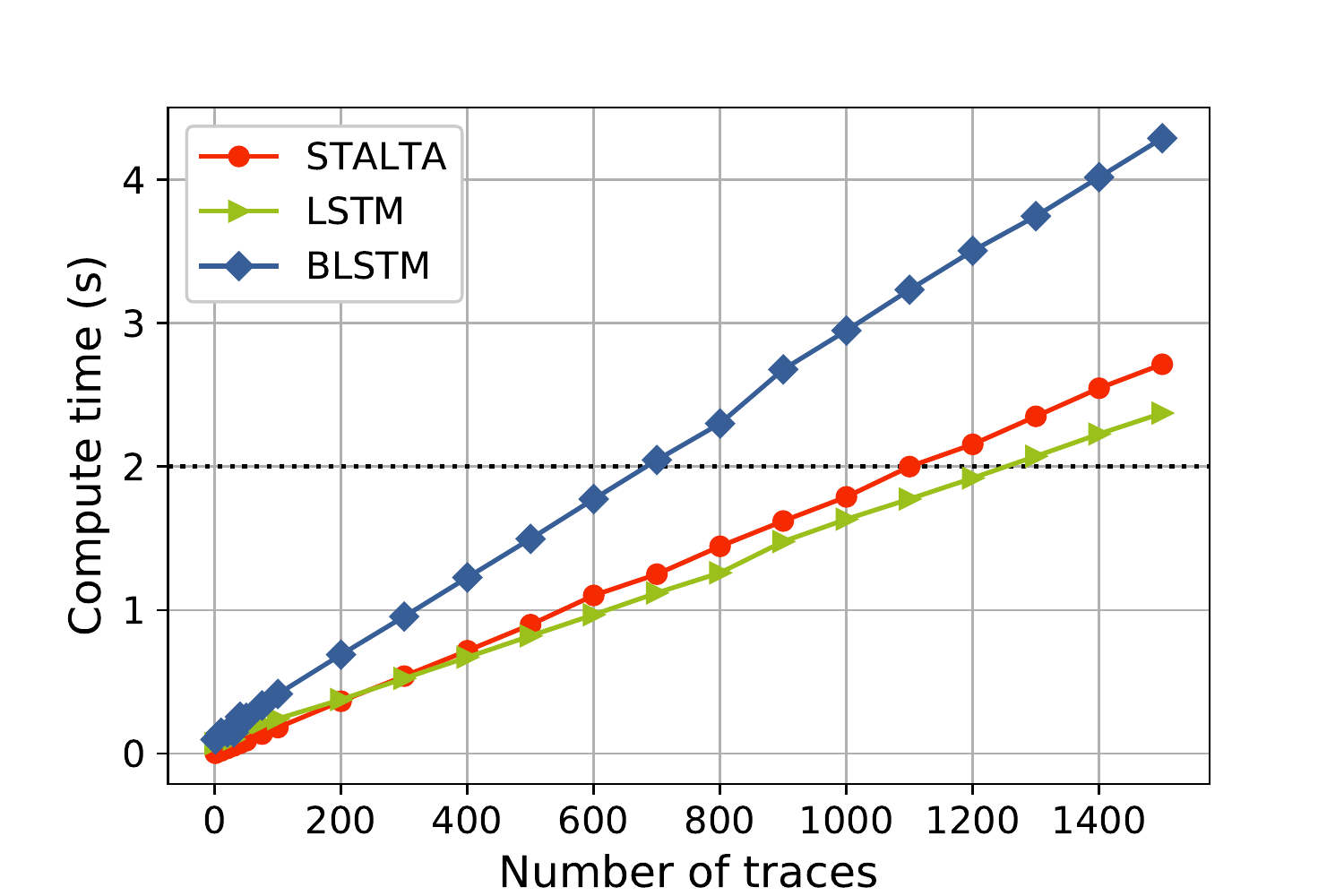}
  \caption{Compute time versus number of traces detection is performed on.}
  \label{fig:comp_times}
\end{figure}

Finally, a real concern surrounding the development of any new event detection procedure is that they can be applied real time. \cite{stork2018} cited this as the reason why new detection approaches are commonly not implemented. Figure \ref{fig:comp_times} illustrates the detection time for a two second window for the three approaches for increasing number of traces\footnote{All compute times are calculated on a Microsoft Azure NC6 virtual machine.}. Whilst the BLSTM has a significantly longer compute time, it can perform detection real time on up to 600 traces on the current setup. As the traces are treated independently then the detection could easily be parallelised over a number of machines. The NN approaches require training prior to their implementation, for 10,000 training samples the training time for the LSTM and BLSTM networks was 45 minutes and one hour respectively. Assuming a training dataset representative of the recording setting, this is a one-off computational cost.

This work has proposed a novel NN-approach to seismic event detection requiring no manual labelling of the data. Whilst it provides a strong competitor to the commonly used STA/LTA event detection procedures, it shares a similar disadvantage in that it is a trace-by-trace approach. Post-processing can be used to attempt to enforce spatial-consistency in detections, for example through the inclusion of a coincidence filter \cite{joswig1993,trnkoczy2012}. However, these approaches are used to filter the results as opposed to being an integral part of the detection procedure. Incorporation of spatial components exponentially increases the number of NN parameters and therefore introduces complications around memory requirements and computation time. \cite{birnie2020tf} detailed a preliminary study investigating the opportunity of using distributed NN training to allow efficient training of large NNs on seismic data. Future work will continue to investigate the optimum methodologies for incorporating spatial information into the training whilst maintaining a realtime detection procedure with reasonable training times.

\section{Conclusion}
This study highlighted how the incorporation of bidirectionality into an LSTM network can improve seismic event detection procedures at low-mid SNRs ($<2.7$). Trained using only synthetic data and benchmarked against the commonly used STA/LTA event detection procedure, the BLSTM approach is shown to significantly increase the number of true detections whilst simultaneously reducing the number of false detections. The potential for the BLSTM application for real time field detection is highlighted by accurate detection of a seismic event from a 3500 sensor PRM array as well as a compute time analysis highlighting the number of traces that can be processed in real time.

\section{Acknowledgements}
The authors would like to thank the Grane license partners Equinor Energy AS, Petoro AS, Vår Energi AS, and ConocoPhillips Skandinavia AS for allowing to present this work. The views and opinions expressed in this abstract are those of the Operator and are not necessarily shared by the license partners.

\newpage

\bibliographystyle{unsrt}  
\bibliography{bibliography}

\end{document}